\documentclass[prb,twocolumn,superscriptaddress]{revtex4}
\usepackage{graphicx}
\usepackage{amsmath}
\usepackage{amsfonts}
\usepackage{amssymb,mathrsfs}
\usepackage{hyperref}

\newcommand{\bs}[1]{\boldsymbol{#1}}

\begin{document}

\title{Vorticity of viscous electronic flow in graphene}

\author{Sven Danz}
\affiliation{\mbox{Institut f\"ur Theorie der kondensierten Materie, Karlsruhe Institute of
Technology, 76128 Karlsruhe, Germany}}
\author{Boris N. Narozhny}
\affiliation{\mbox{Institut f\"ur Theorie der kondensierten Materie, Karlsruhe Institute of
Technology, 76128 Karlsruhe, Germany}}
\affiliation{National Research Nuclear University MEPhI (Moscow Engineering Physics Institute),
  115409 Moscow, Russia}

\date{\today}

\begin{abstract}
In ultra-pure materials electrons may exhibit a collective motion
similar to the hydrodynamic flow of a viscous fluid, the phenomenon
with far reaching consequences in a wide range of many body systems
from black holes to high-temperature superconductivity. Yet the
definitive detection of this intriguing behavior remains
elusive. Until recently, experimental techniques for observing
hydrodynamic behavior in solids were based on measuring macroscopic
transport properties, such as the ``nonlocal'' (or ``vicinity'')
resistance, which may allow alternative interpretation. Earlier this
year two breakthrough experiments demonstrated two distinct imaging
techniques making it possible to ``observe'' the electronic flow
directly. We demonstrate that a hydrodynamic flow in a long Hall bar
(in the absence of magnetic field) exhibits a nontrivial vortex
structure accompanied by a sign-alternating nonlocal resistance. An
experimental observation of such unique flow pattern could serve a
definitive proof of electronic hydrodynamics.
\end{abstract}

\maketitle

Traditional fluid mechanics \cite{dau6} describes long-distance
properties of conventional (e.g., water, oil, etc.) and quantum (e.g.,
$^3$He) fluids equally well \cite{chai,wolf}. The key feature uniting
these diverse many body systems is the short-ranged or collision-like
nature of interactions between the constituent particles that conserve
momentum. In the simplest case of a dilute gas (e.g., air) the
hydrodynamic equations can be derived from the kinetic theory
\cite{dau10}. The resulting theory is however more general than the
derivation: the hydrodynamic equations provide a universal description
applicable to any system within the same symmetry class.

The usual theory of the electron transport in solids also relies on
the kinetic theory \cite{ziman}, but with momentum-relaxing scattering
processes typically dominating over the momentum-conserving
electron-electron interaction. Unlike the conventional fluids,
electrons in solids exist in the environment created by a crystal
lattice where scattering off either lattice imperfections (or
``disorder'') or lattice vibrations (``phonons'') does not conserve
momentum. At length scales exceeding the mean free path
$\ell_{\rm dis}$ (or $\ell_{\rm e-ph}$) the electrons exhibit
diffusive motion \cite{chai,ziman}. In relatively small, mesoscopic
samples of the size smaller than (or comparable to) the mean free
path, ${L\lesssim\ell_{\rm{dis}},\ell_{\rm e-ph}}$, the electrons can
cross the sample ballistically \cite{imry,sulp}.

In contrast, should one manage to fabricate a sample, where the
momentum-conserving electron-electron interaction were the dominant
scattering process, the macroscopic flow of electrons would be
hydrodynamic \cite{rev,luc,gurzhi}. Envisioned by Gurzhi \cite{gurzhi}
long time ago, this idea got traction only with the emergence of
ultra-pure materials
\cite{imm,imh,gal,geim4,geim3,ihn,goo,mac,kim1,geim1}. Yet, while it
has been established that transport properties of such systems deviate
significantly from the traditional expectation \cite{ziman}, the
``hydrodynamic'' interpretation of the observed results may still be
considered as controversial. In particular, the nonlocal resistance (a
tool often used to uncover hydrodynamic transport
\cite{geim4,geim3,geim1}) has been extensively studied in the context
of ballistic transport in multiterminal measurements in the presence
\cite{geimnl2,skocpol,geimnl1} or absence
\cite{nlgold,vwees1,roukes,vonklit} of an external magnetic
field. Thus the hydrodynamic flow can be distinguished from a
ballistic one \cite{fl18} by employing additional measurements: the
nonlocal resistance should be supplemented by measurements of its
temperature dependence \cite{geim3} and even imaging of the Poiseuille
velocity profile has to be supplemented by measurements of additional
quantities \cite{imm}.

Recently two breakthrough experiments demonstrated two distinct
imaging techniques allowing to ``observe'' the electronic flow
directly \cite{imm,imh,sulp}. This makes it possible to study a
distinct feature of the hydrodynamic flow -- vorticity -- in
laboratory experiments. The hydrodynamic vortices (otherwise known as
eddies or whirlpools) were previously invoked to explain the observed
negative nonlocal resistance in graphene \cite{geim1,fl0,pol15}, but
so far have not been directly imaged. In this paper, we solve the
unconventional hydrodynamic equations in graphene \cite{hydro1,me1} in
a realistic Hall bar geometry (away from neutrality) and demonstrate
that the electronic flow forms a unique pattern exhibiting multiple
vortices. Alternation in spinning direction of these vortices results
in a sign-alternating nonlocal resistance \cite{lev19}. These features
could be used to distinguish the hydrodynamic flow (to which they are
unique) from ballistic propagation.

Within linear response and in the case of stationary flow, the
Navier-Stokes equation (\ref{eq1g}) in graphene away from charge
neutrality (i.e., the system considered in the imaging experiments of
Refs.~\onlinecite{sulp,imm}) takes the form \cite{me1,luc,fl0}
\begin{equation}
\label{eq0}
v_g^2 \bs{\nabla} P
=
v_g^2 
\left[
\eta \Delta\bs{u}
+
en\bs{E}
\right]
-
\frac{\mu n\bs{u}}{\tau_{{\rm dis}}},
\end{equation}
where $\eta$ is the shear viscosity \cite{geim1,geim4,me2}, $n$ is the
carrier density, $\bs{E}$ is the electric field, $v_g$ is the Fermi
velocity, $P$ is the thermodynamic pressure, $\mu$ is the chemical
potential, and $\tau_{\rm{dis}}$ is the disorder mean free time, see
Appendix~\ref{hydro} for details. At high enough densities, the
electric current is proportional to the hydrodynamic velocity,
$\bs{J}=en\bs{u}$. In the rest of the paper we analyze this equation
numerically and establish the local flow patterns in experimentally
relevant geometries.

\begin{figure}[t]
\centerline{\includegraphics[width=0.8\columnwidth]{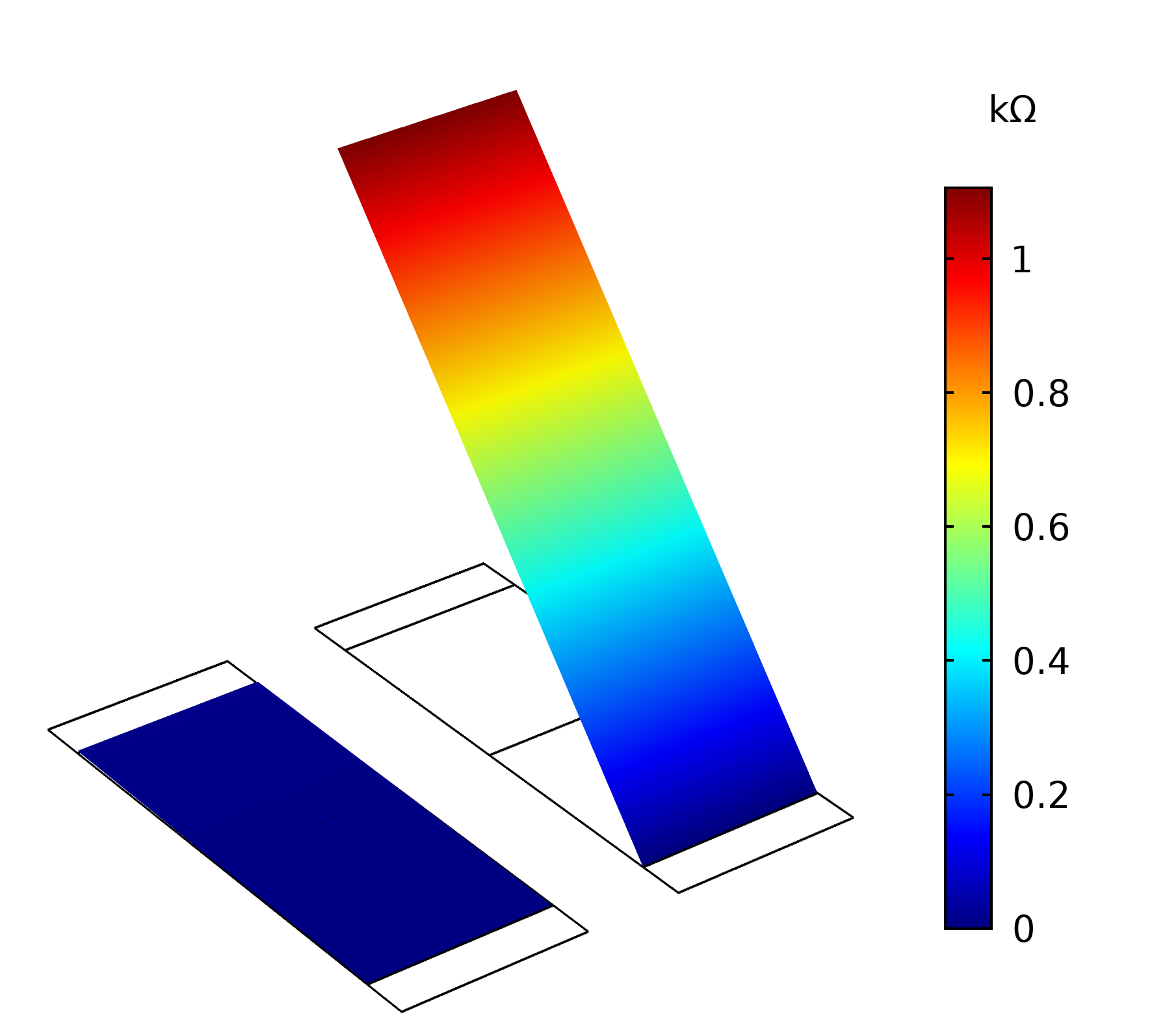}
}
\caption{Ballistic-to-diffusive crossover in inviscid electronic flow
  in doped graphene (cf. Fig.~2 of Ref.~\onlinecite{sulp}). The two
  plots show the electrochemical potential $\Phi$ normalized by the
  driving current $I$ for $\ell_{\rm{dis}}=26\,\mu$m (ballistic
  propagation; left) and $\ell_{\rm{dis}}=0.15\,\mu$m (diffusive flow;
  right). The calculations were performed at $T=4\,$K,
  $n=10^{12}\,$cm$^{-2}$, and $L=25\,\mu$m. The left plot does not
  show the step-like contact voltage drops \cite{sulp,fal19} since the
  contact resistance \cite{alf} was not included in the calculation.}
\label{fig1:baldif}
\end{figure}

To verify our numerical methods (see Appendix~\ref{method}), we first
consider the hydrodynamic equation (\ref{eq0}) in the absence of
viscosity, $\eta=0$. In this case, Eq.~(\ref{eq0}) describes the usual
Ohmic flow of electrons. By varying the ratio of the mean free path,
$\ell_{\rm{dis}}=v_g\tau_{\rm{dis}}$, to the sample length $L$ within
the two-terminal measurement scheme we observe the
ballistic-to-diffusive crossover \cite{sulp}, see
Fig.~\ref{fig1:baldif}. Combining the pressure gradient and the
electric field into the gradient of the electrochemical potential
\cite{fl0,msw}, we follow the change of the latter along the
rectangular sample placed between the two contacts. Choosing the mean
free path to exceed the sample length ($\ell_{\rm{dis}}=26\,\mu$m, see
Ref.~\onlinecite{sulp}) we observe the almost flat profile of the
electrochemical potential across the sample indicative of ballistic
propagation (cf. the effect of electric field expulsion \cite{fal19}),
see the left panel in Fig.~\ref{fig1:baldif}. In contrast, a system
with a smaller mean free path (${\ell_{\rm{dis}}=0.15\,\mu}$m)
exhibits a linear variation of the electrochemical potential typical
of the diffusive transport, see the left panel in
Fig.~\ref{fig1:baldif}. Not surprisingly, this behavior fully agrees
with experimental observations (see Fig.~2 of Ref.~\onlinecite{sulp}).

Viscous flow in the same two-terminal geometry is expected to exhibit
a Poiseuille-like velocity profile \cite{imm,luc,fl2,mr2}. The precise
form of the profile is determined by the interplay of the viscosity,
disorder mean free time, and boundary conditions. The parabolic
profile \cite{dau6} corresponds to a pure liquid
(${\tau_{\rm{dis}}\rightarrow\infty}$) with the ``no-slip'' boundary
conditions (${\bs{u}=0}$ at the sample edges). The latter were
recently shown to be inapplicable in graphene \cite{ks19}; instead,
one should apply a more general boundary condition, Eq.~(\ref{sbc}),
characterized by the temperature-dependent ``slip length''. This
quantity was recently measured in graphene \cite{imm} to be
${\zeta\approx500}\,$nm at ${T=75}\,$K.

\begin{figure}[t]
\centerline{\includegraphics[width=0.9\columnwidth]{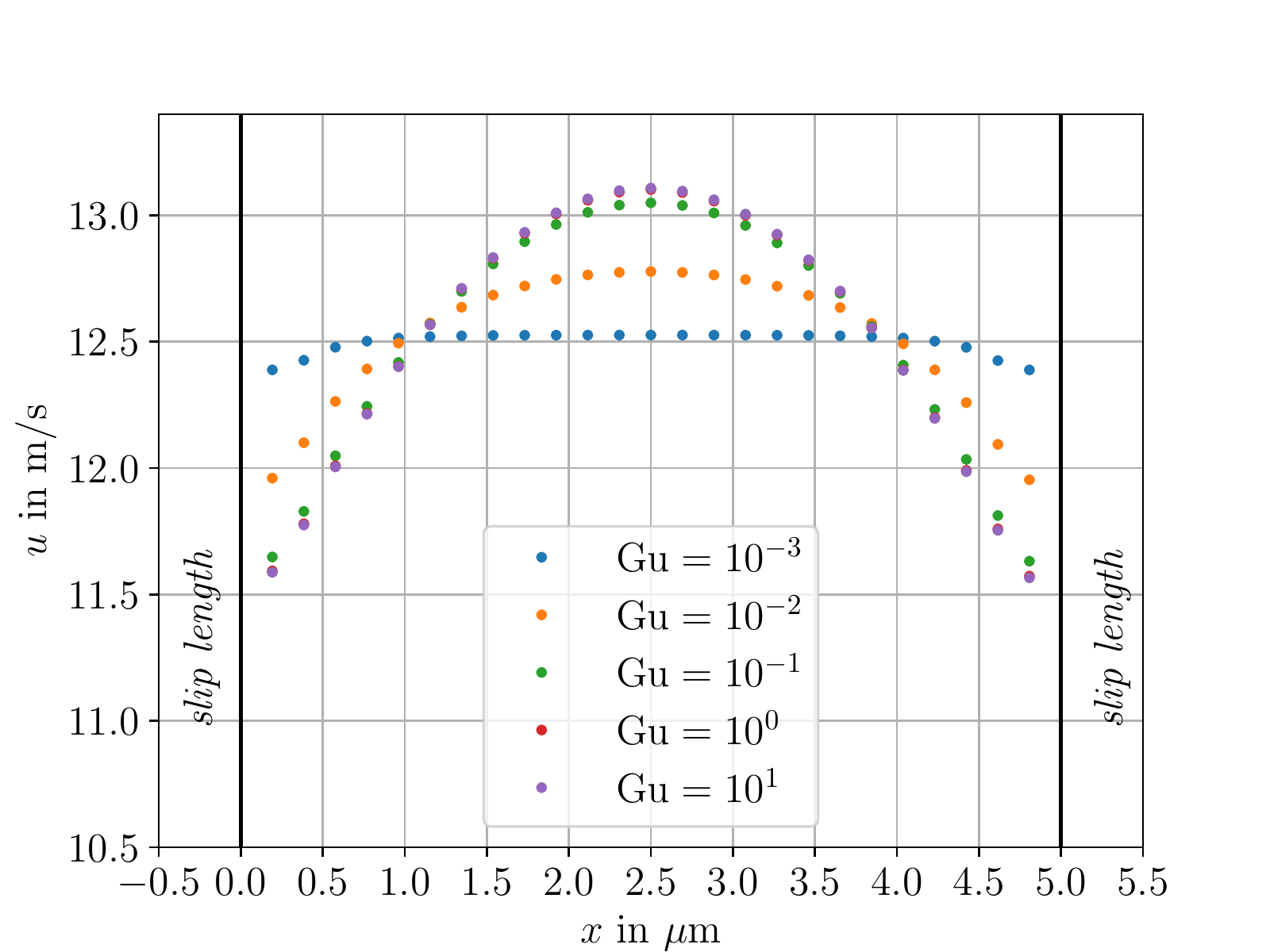}
}
\caption{Flow velocity profile in a viscous electronic flow in a strip
  geometry ($W=5\,\mu$m, $L=10\,\mu$m). The profiles were calculated
  with the ``slip boundary condition'' with the slip length \cite{imm}
  $\zeta=0.5\,\mu$m. The color-coded datasets correspond to different
  values of the Gurzhi number (\ref{gnum}).}
\label{fig2:upo}
\end{figure}

To describe the mutual effect of disorder and viscosity on the
electronic flow we introduce a ``Gurzhi number''
\begin{equation}
\label{gnum}
{\rm Gu} = \frac{\nu\tau_{\rm{dis}}}{l^2},
\qquad
\nu = \frac{v_g^2\eta}{{\cal W}},
\end{equation}
where $\nu$ is the kinematic viscosity \cite{dau6,geim1,geim4,me2},
${\cal W}$ is the enthalpy [see Eq.~(\ref{eqsta})] and $l$ is a
typical length scale in the problem (e.g., the width of the graphene
strip) in analogy to the standard composition of the Reynolds number
\cite{dau6}. For small values of the Gurzhi number the electrons
propagate ballistically with a flat velocity profile across the bulk
of the sample, while large values of ${\rm Gu}$ correspond to a
Poiseuille-like flow with a nonuniform velocity distribution. This
behavior is illustrated in Fig.~\ref{fig2:upo}, where we show velocity
profiles for various values of ${\rm Gu}$ measured in a strip of width
$W=5\,\mu$m.

\begin{figure*}[t]
\centerline{\includegraphics[width=0.85\textwidth]{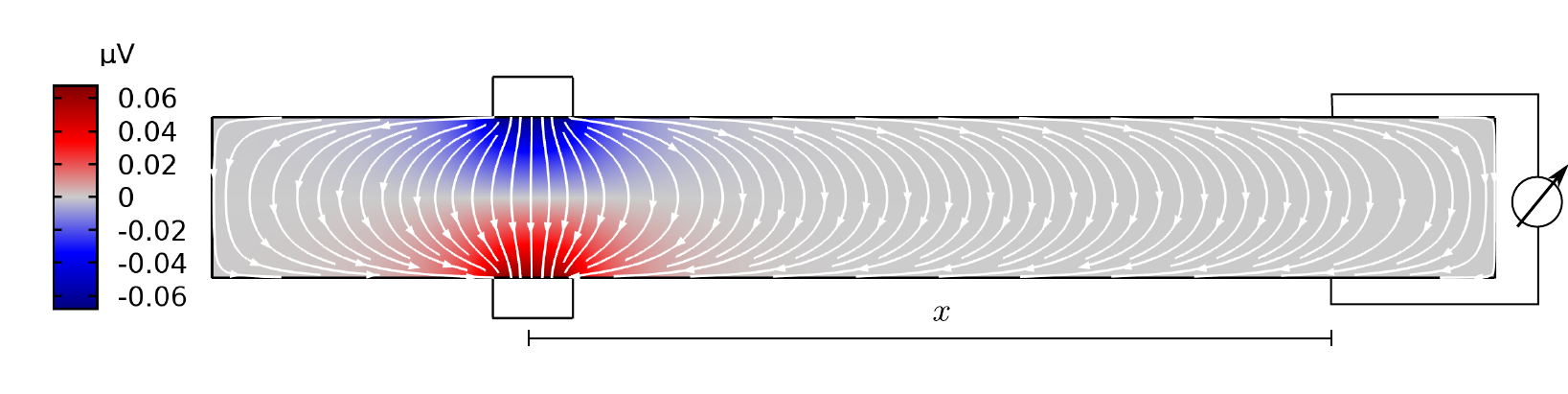}
}
\centerline{\includegraphics[width=0.85\textwidth]{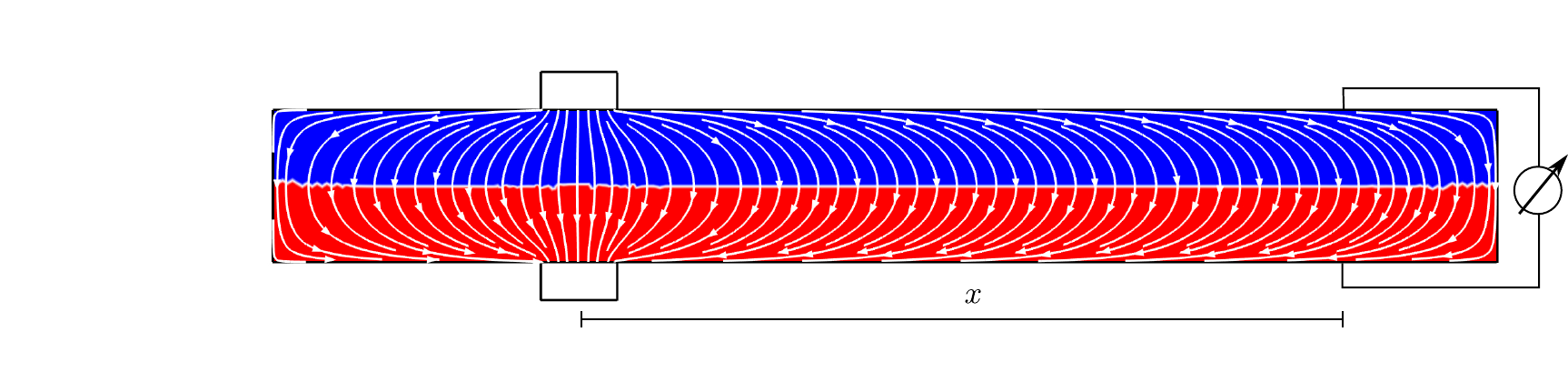}
}
\centerline{\includegraphics[width=0.85\textwidth]{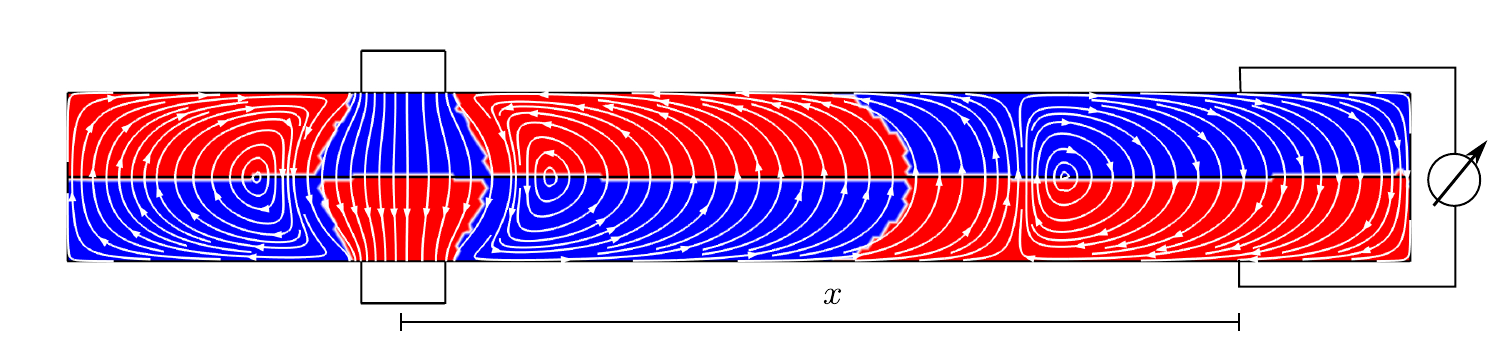}
}
\centerline{\includegraphics[width=0.85\textwidth]{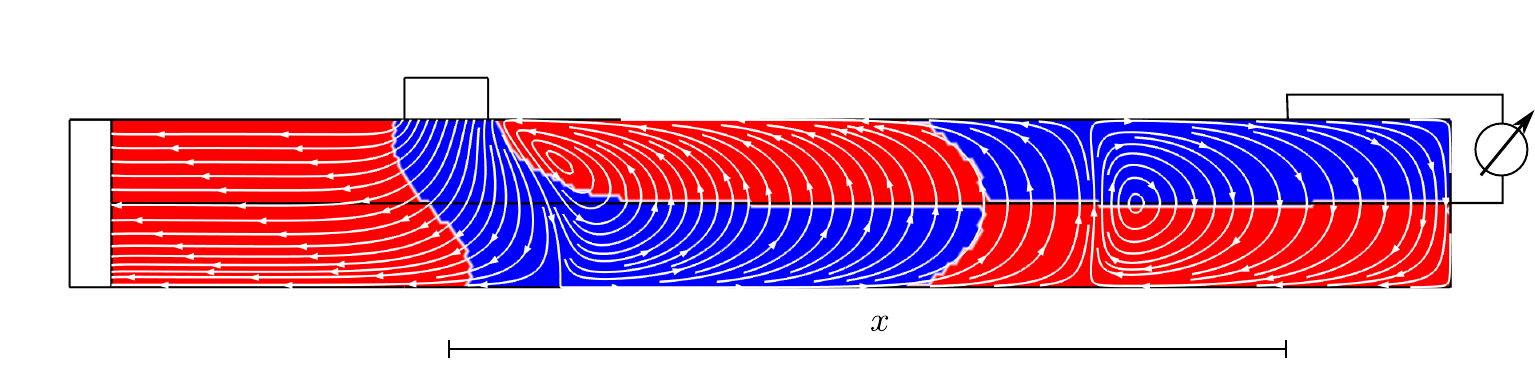}
}
\caption{Hydrodynamic electron flow in a Hall bar. Top two panels:
  Ohmic flow. Two bottom panels: viscous flow exhibiting multiple
  vortices and sign-alternating nonlocal resistance. White rectangles
  indicate the source and drain contacts. The data were calculated
  with the ``slip boundary condition'' Eq.~(\ref{sbc}) with the slip
  length \cite{imm} ${\zeta=0.1}\,\mu$m and for ${W=2}\,\mu$m,
  ${L=16}\,\mu$m, ${T=200}\,$K, ${n=10^{12}}\,$cm$^{-2}$, and ${{\rm
      Gu}=0.175}$ (increasing ${\rm Gu}$ or decreasing $\zeta$ reduces
  the vortex size increasing the number of vortices in the
  sample). The color map in the top panel shows the deviation of the
  electrochemical potential from its median value, while the red and
  blue colors in all other panels indicate the sign of this quantitiy
  (the broken white line separating the red and blue regions indicates
  the median value of the electrochemical potential; its patchiness
  reflects the finite precision of the simulation). The equidistant
  streamlines mask the exponential decay of the flow velocity, see
  Fig.~\ref{fig4:decay1}. }
\label{fig3:flow}
\end{figure*}

\begin{figure}[t]
\centerline{\includegraphics[width=0.9\columnwidth]{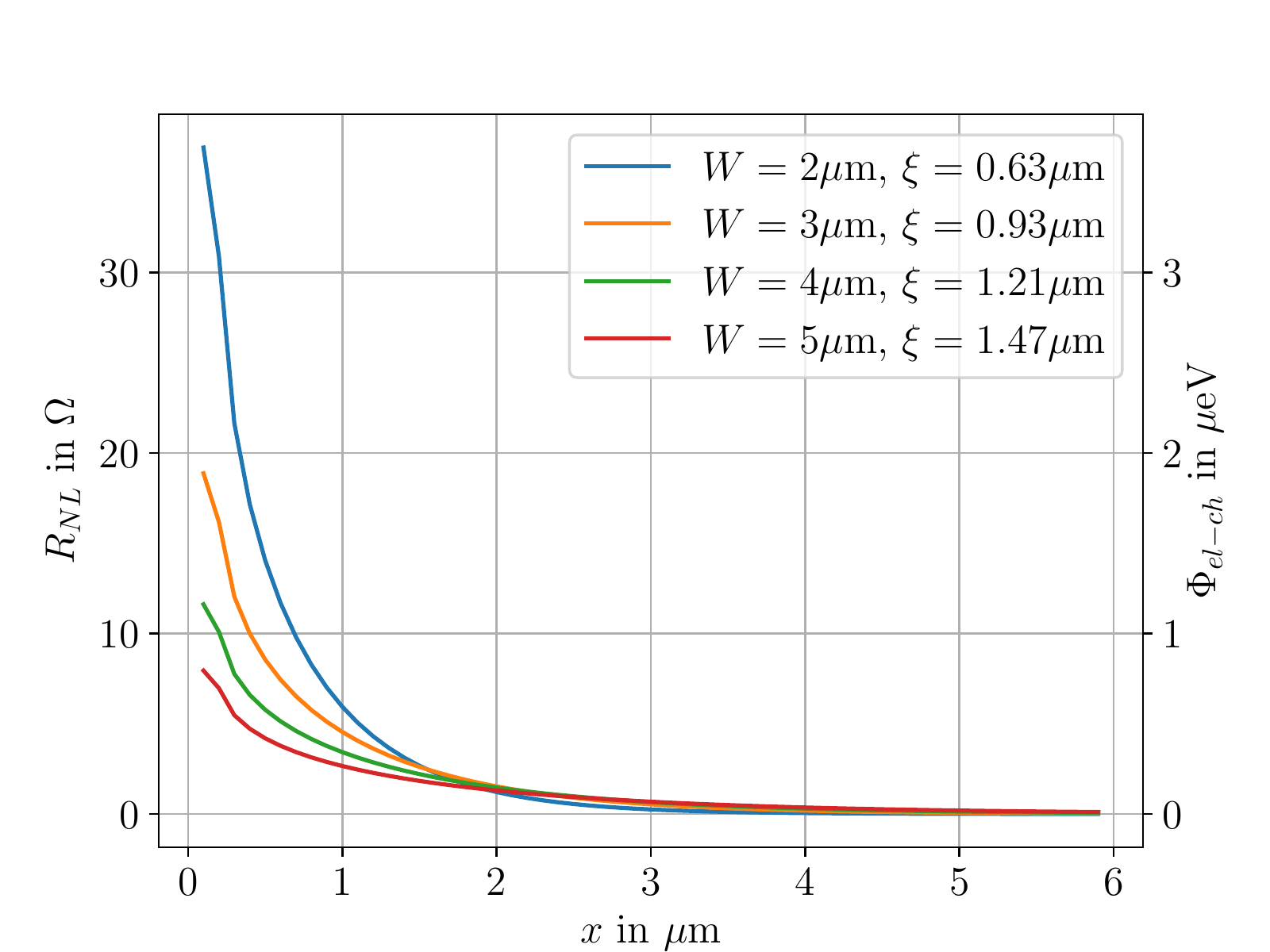}
}
\caption{Signal decay in the Ohmic flow (shown in the top panel in
  Fig.~\ref{fig3:flow}). The nonlocal resistance measured at a
  distance $x$ away from the source and drain contacts decays
  exponentially, $R_{NL}\propto\exp(-x/\xi)$. The obtained values of
  $\xi$ for different widths $W$ (shown in the legend) roughly
  agree with the estimate $\xi=W/\pi$ based on the van der Pauw method
  \cite{nlgold,nlr}.}
\label{fig4:decay1}
\end{figure}

The nonlocal effects in the electronic flow in graphene are typically
studied in a Hall bar geometry \cite{imm,geim1,geim3}. The inviscid,
Ohmic current is expected to flow directly from the source to drain
contacts \cite{fl1,fl2}, while exponentially decaying into the bulk of
the Hall bar. In long mesoscopic systems (i.e. with $L>W$) such decay
is governed by the sample geometry rather than momentum relaxation as
follows from the van der Pauw theorem \cite{pauw,nlgold,nlr}. Our
numerical solution to Eq.~(\ref{eq0}) demonstrates exactly this
behavior, see Figs.~\ref{fig3:flow} and \ref{fig4:decay1}. The top
panel in Fig.~\ref{fig3:flow} shows the flow pattern of the Ohmic flow
across a Hall bar. Measuring the voltage drop (i.e., calculating the
chemical potential difference) at a distance $x$ away from the source
and drain contacts yields an exponentially decaying signal as
illustrated in Fig.~\ref{fig4:decay1}. We have verified that the decay
length $\xi$ (defined in the figure caption) is mostly determined by
the width of the Hall bar in agreement with the standard expectation
based on the van der Pauw method \cite{nlgold,nlr} and remains largely
insensitive to a variation of either $L$ or $\ell_{\rm dis}$.

The color map in the top panel in Fig.~\ref{fig3:flow} shows the
electrochemical potential distribution in the Hall bar. Away from the
source and drain contacts, the electrochemical potential decays
exponentially maintaining its sign along the Hall bar. To illustrate
this point, the second panel in Fig.~\ref{fig3:flow} shows the sign of
the deviation of the electrochemical potential from its median
value. The nonlocal resistance (at a distance $x$ away from the source
and drain contacts) is exponentially small, see
Fig.~\ref{fig4:decay1}, and positive.

Viscous flow in the Hall bar geometry is accompanied by formation of
vortices \cite{geim1,fl1,fl0,pol15} (also referred to as eddies or
whirlpools). In traditional hydrodynamics \cite{dau6}, vorticity
appears already at the level of the Euler equation (describing ideal,
inviscid flows), while viscosity tends to suppress it. Laminar flows
(with small Reynolds number) tend to be vortex-free far away from flow
boundaries or obstacles with the latter being ``responsible'' for
nontrivial flow patterns (e.g., in the laminar wake \cite{dau6}).

Now, electron motion in mesoscopic samples almost always takes place
in proximity of the sample boundaries and hence the appearance of
nontrivial vortex structure is to be expected. Earlier measurements
aiming at detection hydrodynamic flows of electrons in graphene
\cite{geim1} used the nonlocal resistance $R_{NL}$ as the tool to
uncover nontrivial flow patterns: a change of direction of the
electronic flow may lead to a sign change in $R_{NL}$. The negative
nonlocal resistance observed in these experiments was then interpreted
\cite{fl0,fl1,pol15} as a sign of vortex formation indicating a
viscous flow. Later on it became clear that negative nonlocal
resistance may appear in multiterminal measurements on ballistic (and
perhaps even diffusive \cite{nlgold}) electronic systems
\cite{geim3,imm,fl18} such that additional measurements were necessary
to ensure that one observes indeed the hydrodynamic behavior.

\begin{figure}[t]
\centerline{\includegraphics[width=0.9\columnwidth]{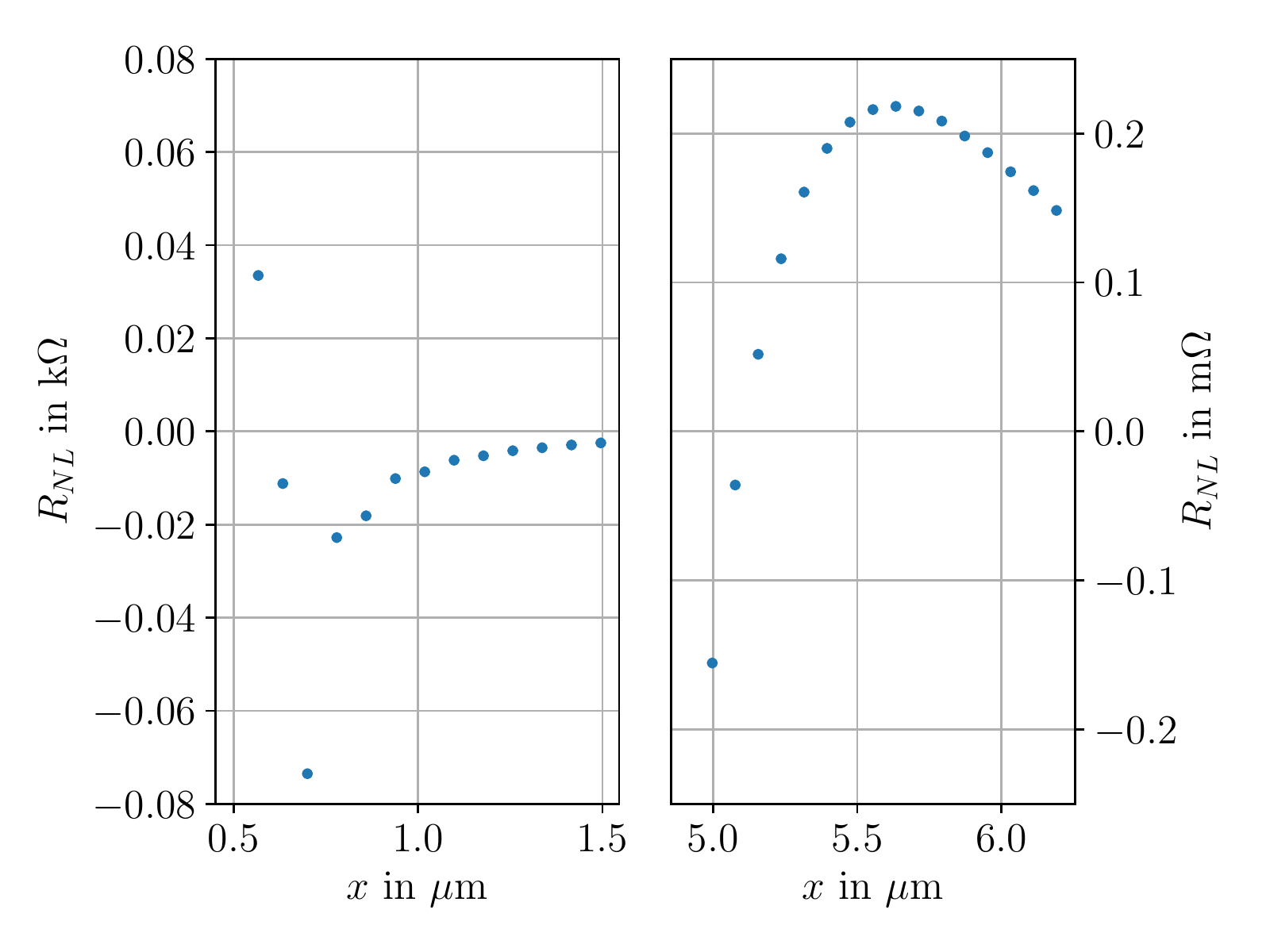}
}
\caption{Sign-alternating nonlocal resistance in viscous electronic
  flow (see the two bottom panels in Fig.~\ref{fig3:flow}). While
  still decaying exponentially, the nonlocal resistance measured in a
  viscous flow exhibiting multiple vortices changes sign as a function
  of the distance $x$ away from the source and drain contacts. Left
  panel: the first sign change in $R_{NL}$ due to the first vortex
  observed in Ref.~\onlinecite{geim1} and calculated in
  Ref.~\onlinecite{fl0}. Right panel: the second sign change in
  $R_{NL}$ due to the second vortex with vorticity opposite to that of
  the first vortex. While the values of $R_{NL}$ are exponentially
  smaller, they remain in the measurable m$\Omega$ range.}
\label{fig5:nlr}
\end{figure}

We have solved the hydrodynamic equation (\ref{eq0}) with the slip
boundary conditions \cite{imm,ks19} in a somewhat longer (in
comparison to the experimental samples of
Refs.~\onlinecite{geim1,geim3,geim4}), gated
\cite{geim1,geim3,ash,mr1} Hall bar. The results of our calculations
are shown in the two bottom panels in Fig.~\ref{fig3:flow} and in
Fig.~\ref{fig5:nlr}. The third panel in Fig.~\ref{fig3:flow} shows the
flow pattern in the ``theoretical'' geometry considered in
Refs.~\onlinecite{fl0,fl1} with the source and drain contacts located
on the opposite sides of the Hall bar. The bottom panel in
Fig.~\ref{fig3:flow} shows the current distribution in the
``experimental'' geometry of Ref.~\onlinecite{geim1} (see also
Ref.~\onlinecite{pol15}). In both cases, the flow pattern on the right
of the source contact (where we calculate the nonlocal resistance) is
rather similar. The first vortex appearing to the right of the source
contact is located at a distance roughly equal to the width of the
Hall bar \cite{fl0} and is somewhat asymmetric stretching more in the
direction away from the contact. In a shorter Hall bar (as in the left
side of the third panel in Fig.~\ref{fig3:flow}) there would be space
for only one vortex, but in a longer system the second vortex
appears. The size of the vortices and the exact location of the second
vortex is determined by the interplay of the sample geometry, slip
length (see Appendix~\ref{method}), and disorder mean free path (we
chose ${{\rm Gu}=0.175}$ on the basis of experimental data
\cite{imm,gal,geim1}; increasing the value of ${\rm Gu}$ or reducing
the slip length reduces the vortex size; this way we may observe more
vortices in the sample). As the second vortex ``spins'' in the
opposite direction to the first one, the resulting nonlocal resistance
exhibit another sign change, see Fig.~\ref{fig5:nlr}. Since the
relation between the flow velocity and electrochemical potential
involves derivatives, the sign change is not aligned with the vortex
boundaries (such that the correspondence between the flow direction
and the sign of $R_{NL}$ is not direct).

Given that the decay of the electronic flow (see
Fig.~\ref{fig4:decay1}) has mostly geometric origin
\cite{pauw,nlgold,nlr}, it is not surprising that the hydrodynamic
flow decays in a similar fashion. As a result, the nonlocal resistance
after the second sign change (see the right panel in
Fig.~\ref{fig5:nlr}) is several orders of magnitude smaller than the
absolute value of the negative $R_{NL}$ after the first sign
change. Nevertheless, using realistic parameters (with the values
borrowed from the experiments of
Refs.~\onlinecite{geim1,geim3,gal,imm}) we obtain values of order
m$\Omega$ (measurable in a modern laboratory).

To conclude, we have solved the hydrodynamic equations in doped
graphene in experimentally relevant Hall bar geometry. We have mapped
the nontrivial vortex pattern of the viscous electronic flow and
calculated the corresponding sign-alternating nonlocal resistance. In
contrast to Ref.~\onlinecite{lev19}, this behavior is not related to
quasiparticle recombination (which is only important in close
proximity to the Dirac point) and magnetic field (see also
Ref~\onlinecite{sven2}). At charge neutrality, the electric current
density in the absence of magnetic field is decoupled from the
hydrodynamic velocity, see Eq.~(\ref{djdef}). As a result, transport
properties of neutral graphene are more complicated
\cite{imh,sven2,hydro0,mr1,mr3} and require further investigation. We
believe that the obtained flow patterns are unlikely to occur in
ballistic electronic systems. We expect that an experimental
observation of multiple vortices with the help of modern imaging
techniques of Refs.~\onlinecite{sulp,imm,imh} combined with the
measurement of the sign-alternating nonlocal resistance will uniquely
identify the hydrodynamic behavior of charge carriers in graphene, as
well as any other electronic system that is pure enough to support the
``hydrodynamic temperature window'' [based on the universality of the
  hydrodynamic equation (\ref{eq0}) which -- unlike Eq.~(\ref{eq1g})
  -- is no longer specific to Dirac fermions in graphene].

\section*{Acknowledgments}

The authors wish to thank I.V. Gornyi, A.D. Mirlin, J. Schmalian,
J.A. Sulpizio, M. Sch\"utt, and A. Shnirman for fruitful
discussions. This work was supported by the German Research Foundation
DFG within FLAG-ERA Joint Transnational Call (Project GRANSPORT), by
the European Commission under the EU Horizon 2020 MSCA-RISE-2019
program (Project 873028 HYDROTRONICS), and the MEPhI Academic
Excellence Project, Contract No. 02.a03.21.0005.

\appendix
\section{}
\label{hydro}

The hydrodynamic description of electronic transport in graphene was
discussed in detail in Refs.~\onlinecite{hydro1,me1} (see also
Refs.~\onlinecite{rev,luc}). In the absence of the magnetic field the
generalized Navier-Stokes equation in graphene is
\begin{eqnarray}
\label{eq1g}
&&
\!\!\!\!\!\!
{\cal W}(\partial_t+\bs{u}\!\cdot\!\bs{\nabla})\bs{u}
+
v_g^2 \bs{\nabla} P
+
\bs{u} \partial_t P 
+
e(\bs{E}\!\cdot\!\bs{j})\bs{u} 
=
\\
&&
\nonumber\\
&&
\qquad\qquad\qquad\qquad\qquad
=
v_g^2 
\left[
\eta \Delta\bs{u}
+
en\bs{E}
\right]
-
\bs{j}_E/\tau_{{\rm dis}},
\nonumber
\end{eqnarray}
where $\eta$ is the shear viscosity \cite{geim1,geim4,me2}, $n$ is the
carrier density, $\bs{E}$ is the electric field, $v_g$ is the Fermi
velocity in graphene, and $\tau_{{\rm dis}}$ is the disorder mean free
time. The enthalpy ${\cal W}$ and pressure $P$ are related to the energy
density $n_E$ in graphene by the ``equation of state''
\cite{hydro1,me1}
\begin{equation}
\label{eqsta}
{\cal W}=n_{E}+P = 3n_{E}(2\!+\!u^2/v_g^2)^{-1}.
\end{equation}
The carrier and energy currents ($\bs{j}$ and $\bs{j}_E$,
respectively) are related to the hydrodynamic
velocity $\bs{u}$ as
\begin{equation}
\label{djdef}
\bs{j} = n \bs{u} + \delta\bs{j},
\qquad
\bs{j}_{E} = {\cal W}\bs{u} 
,
\end{equation}
where $\delta\bs{j}$ is the dissipative correction \cite{luc,me3}. The
full electric current is given by $\bs{J}=e\bs{j}$. The hydrodynamic
theory is completed by the continuity equations
\begin{subequations}
\label{ces}
\begin{equation}
\label{cen1}
\partial_t n + \bs{\nabla}_{\bs{r}}\!\cdot\!\bs{j} = 0,
\end{equation}
\begin{equation}
\label{cene1}
\partial_t n_E + \bs{\nabla}_{\bs{r}}\!\cdot\!\bs{j}_E = e \bs{E}\cdot\bs{j},
\end{equation}
\end{subequations}
where the last term describes Joule heat. The continuity equation
(\ref{cen1}) is valid in any electronic system, while
Eq.~(\ref{cene1}) neglects possible energy losses due to coupling to
collective excitations (e.g., phonons, plasmons, etc.).

Consider a graphene sample away from charge neutrality, i.e. the
system considered in the imaging experiments of
Refs.~\onlinecite{sulp,imm}. In this case the densities are determined
by the chemical potential $\mu$ 
\begin{equation}
\label{flr}
n(\mu\gg{T})=\frac{\mu^2}{\pi v_g^2},
\;\;
n_E(\mu\gg{T})=\frac{2\mu^3}{3\pi v_g^2},
\;\;
\delta\bs{j}=0.
\end{equation}
Similarly, the mean free time assumes the value
$\tau_{\rm{dis}}(\mu)$. Within linear response and in the case of
stationary flow, the Navier-Stokes equation (\ref{eq1g}) takes the
form (\ref{eq0})
\begin{equation}
\label{eq0a}
v_g^2 \bs{\nabla} P
=
v_g^2 
\left[
\eta \Delta\bs{u}
+
en\bs{E}
\right]
-
\mu n\bs{u}/\tau_{{\rm dis}}.
\end{equation}
In the rest of the paper we analyze this equation numerically and
establish the local flow patterns in experimentally relevant
geometries.

In the textbook case of a uniform current \cite{ziman,luc,me1} we
recover the standard Drude formula
\begin{equation}
\label{sdrude}
\bs{J}\!=\!en\bs{u}
\!=\!
(e^2/\pi)\mu\tau_{{\rm dis}} \bs{E}
\;\;\Rightarrow\;\;
\sigma=(e^2/\pi)\mu\tau_{{\rm dis}}.
\end{equation}
Relating the pressure to the carrier density
\begin{equation}
\label{pgrad}
\bs{\nabla}P = \frac{n}{\rho(0)}\bs{\nabla}n, 
\qquad
\rho(0)=\frac{\partial n}{\partial\mu},
\end{equation}
where $\rho(0)$ is the density of states (DoS), and taking into
account the local charge density variations in Eq.~(\ref{eq0}), one
arrives at the usual description of the diffusive transport
\begin{equation}
\label{diff}
\bs{J} = (e^2/\pi)\mu\tau_{{\rm dis}} \bs{E} - eD\bs{\nabla}n,
\qquad
D=v_g^2\tau_{\rm{dis}}/2,
\end{equation}
where $D$ is the diffusion coefficient (see Ref.~\onlinecite{dau10};
for a simplified discussion in a similar context see
Ref.~\onlinecite{df2}).

\section{}
\label{method}

Our numerical technique of choice for an analytically intractable
problem [e.g., the Navier-Stokes equation (\ref{eq1g}) in a realistic
  geometry] is the finite element method. With this method we can
calculate the velocity $\bs{u}$ and carrier density $n$ up to a finite
accuracy, which is given by the number of elements in which we split
the system. A solution can be approximated by an auxiliary function
\[
u_h(x)=\sum_{i=0}^N u_i \varphi_i(x),
\]
interpolating between the predetermined points separating the
elements. The test function $\varphi_i(x)$ can be chosen arbitrarily
provided it satisfies
\begin{align*}
\varphi_i(x)=\begin{cases}
1, & \text{if $x=x_i$}\\
0, & \text{if $x\neq x_i$}
\end{cases}.
\end{align*}
For our purposes, it was sufficient to choose a linear Lagrange
polynomial for the carrier density and a quadratic one for the
hydrodynamic velocity.

This approach can be applied to any linear differential equation of form
\begin{align*}
\mathcal{L} u(x) = g(x),
\end{align*}
which can be solved up to a predetermined accuracy by building its
weak form
\[
\sum_{i=0}^N u_i \int\limits_{\Omega} \mathcal{L} \varphi_i(x) \varphi_j(x) d\Omega 
= \int\limits_{\Omega} g(x)\varphi_j(x) d\Omega.
\]
This yields a linear system of equations with an invertable
${(N+1)\times(N+1)}$ matrix that allows one to find the ${N+1}$ vector
$u_i$ at every point.

The Navier-Stokes equation (\ref{eq1g}) in graphene should be
supplemented by the appropriate boundary conditions. In the case of
the source and drain contacts we apply the finite Dirichlet and zero
flux boundary conditions. For the edges of the graphene Hall bar we
apply the slip boundary condition \cite{ks19}
\begin{equation}
\label{sbc}
\left.u^t_{\alpha}\right|_{\partial\Omega}
-\zeta\bs{n}\cdot\bs{\nabla} u^t_{\alpha}=0,
\end{equation}
where $\bs{n}$ is the unit vector normal to the boundary.

The boundary conditions have a profound effect on the resulting flow
patterns \cite{fl1}. For the commonly used ``no-slip'' boundary
conditions (${\zeta=0}$), the vortices shown in Fig.~\ref{fig3:flow}
exhibit the smallest tail (in the direction away from the source
contact), which is growing with $\zeta$ (assuming all other parameters
remain unchanged). In the opposite limit of ``no-stress'' boundary
conditions (${\zeta\rightarrow\infty}$) the tail length exceeds all
system sizes used in our simulations such that the flow pattern
exhibits a single vortex \cite{fl0,fl1}.

The rectangular contacts shown in Fig.~\ref{fig3:flow} possess sharp
corners. Taken literally, such geometry is plagued with singular
behavior \cite{fl0,wick,gir,kaw}. Given that these singularities have
nothing to do with the observed vorticity \cite{fl1} (in the context
of the Hall effect the corners in real devices are known to be
effectively rounded \cite{gir,kaw}) we smooth out the Dirichlet
boundary condition at the corners of the source contact to obtain
stable flow patterns.

\bibliography{viscosity_refs}

\end{document}